# Low-magnetic-field control of dielectric constant at room temperature realized in $Ba_{0.5}Sr_{1.5}Zn_2Fe_{12}O_{22}$

**Y S Chai**, S H Chun, S Y Haam, **Y S Oh**, **Ingyu Kim** and **Kee Hoon Kim**[1]

*FPRD, Department of Physics and Astronomy, Seoul National University,*

*Seoul 151-742, South Korea*

E-mail: khkim@phya.snu.ac.kr

**Abstract**

We show that room temperature resistivity of $Ba_{0.5}Sr_{1.5}Zn_2Fe_{12}O_{22}$ single crystals increases by more than three orders of magnitude upon being subjected to optimized heat treatments. The increase in the resistivity allows the determination of magnetic field ($H$)-induced ferroelectric phase boundaries up to 310 K through the measurements of dielectric constant at a frequency of 10 MHz. Between 280 and 310 K, the dielectric constant curve shows a peak centered at zero magnetic field and thereafter decreases monotonically up to 0.1 T, exhibiting a magnetodielectric effect of 1.1%. This effect is ascribed to the realization of magnetic field-induced ferroelectricity at an $H$ value of less than 0.1 T near room temperature. Comparison between electric and magnetic phase diagrams in wide temperature- and field-windows suggests that the magnetic field for inducing ferroelectricity has decreased near its helical spin ordering temperature around 315 K due to the reduction of spin anisotropy in $Ba_{0.5}Sr_{1.5}Zn_2Fe_{12}O_{22}$ .

---

[1] Author to whom correspondence should be addressed.

# 1. Introduction

Recently, there has been an increase in the number of researches—both basic and applied researches—on a new class of materials called multiferroics, wherein ferroelectric (FE) and magnetic orders coexist or large magnetoelectric effects are seen [1, 2]. For the practical realization of a multifunctional device, it is required to control the electric polarization ($P$) or dielectric constant ($\varepsilon$) under a small magnetic field ($H$), particularly near room temperature [2]. However, large magnetoelectric or magnetodielectric effects in single phase multiferroic materials have been so far observed mostly at low temperatures [3, 4].

$Zn_2Y$-type hexaferrite $Ba_{0.5}Sr_{1.5}Zn_2Fe_{12}O_{22}$ is a good candidate for improving the current situation because it is predicted that the compound can exhibit $H$-induced ferroelectricity at room temperature under a rather small $H$ of ~0.8 T [5]. Below the Nèel temperature ($T_N$) of 326 K, the compound is known to develop the helical spin ordering, in which the spin moments lie and rotate in the hexagonal $ab$-plane. When $H$ is applied in the $ab$-plane, the compound undergoes several magnetic transitions, among which the so-called intermediate-III phase is found to have a finite value of $P$. Furthermore, on the basis of the $H$-dependent magnetization $M(H)$ measurements, the intermediate-III phase, i.e., the FE phase, has been suggested to exist up to $T_N$ = 326 K, while both $P(H)$ and $\varepsilon(H)$ measurements have provided evidence for the occurrence of ferroelectricity only up to 130 K. Above 130 K, those electrical measurements could not be carried out to confirm the occurrence of ferroelectricity due to the low resistivity of the specimen [5]. Thus, it is necessary to increase the resistivity of this hexaferrite system in order to corroborate the occurrence of $H$-induced ferroelectricity through electrical measurements and to observe magnetoelectric coupling around room temperature.

In this paper, we report the effect of heat treatments on both resistivity ($\rho$) and FE phase boundaries of $Ba_{0.5}Sr_{1.5}Zn_2Fe_{12}O_{22}$ single crystals. By optimizing the heat treatment conditions, we achieve an increase in the resistivity by more than three orders of magnitude at 300 K. This increase enables us to determine the magnetic field-induced FE phase boundaries up to 310 K.

2. **Experimental details**

Single crystals of $Ba_{0.5}Sr_{1.5}Zn_2Fe_{12}O_{22}$ were grown from $Na_2O/Fe_2O_3$ flux in air. After being melted at 1420°C in a Pt crucible, the flux mixture was subjected to several thermal cycling to avoid impurity phase [6] and cooled to room temperature at a rate of 50°C/h. X-ray diffraction of the as-grown crystals showed lattice parameters consistent with the literature values [6]. The as-grown crystals were annealed in a flowing $O_2$ atmosphere at 900°C for 2, 8, and 14 days and then cooled down at three different rates—*quench*, *intermediate*, and *slow*. The rate *quench* refers to the case of quenching from 900°C to room temperature under air, *intermediate* refers to the process of cooling to room temperature by turning off the electric power at 900°C, and *slow* refers to a slow cooling with a rate of 50°C/h. The average cooling rate of *intermediate* was 800°C/h. Through this various treatment conditions, we found the largest resistivity could be obtained at room temperature in a sample treated with 8 day-annealing and *slow* cooling, which will be called as the '*most insulating*'-sample below (See, section 3 for resistivity data).

A standard four-probe method was used to obtain the temperature-dependent $\rho$ of all the as-grown crystals in the *ab*-plane. Due to the high resistance of the heat-treated crystals, a two-probe method was employed to measure their $\rho(T)$. The samples for dielectric measurements are polished into wide and thin plates in a way that electric field (*E*) and *H* are in the *ab*-plane and perpendicular to each other (as depicted in figure 3(c) inset). A typical dimension of the sample used is $2\times0.3\times1$ mm$^3$. Dielectric constant was measured with an LCR meter (Hewlett-Packard 4275A) at a frequency of 10 MHz. In order to trace ferroelectric phase boundary by measuring $\varepsilon(H)$ peaks under minimal influence of dielectric loss, we have measured complex dielectric constant of each crystal, $\varepsilon + i\varepsilon_2$, in a broad frequency range to choose the frequency with the minimum loss. From this investigation, we have found that the dielectric loss of most samples including the '*most-insulating*' one becomes more or less minimum, particularly at the frequency of 10 MHz near room temperature.

We have also measured the magnetoelectric current ($J_{ME}$) by using an electrometer (Keithley 6517) while sweeping magnetic field at a rate of 20 Oe/sec at each temperature. Before the $J_{ME}$ measurement, we have poled each sample with $E$=250 kV/m in its paraelectric state (at $\mu_0H$=4 T), and then adjusted $\mu_0H$=1 T to drive the sample into the ferroelectric state. After these processes, $E$ is removed and $J_{ME}$ is measured while sweeping $H$ up or down.

3. **Results and discussion**

Figure 1(a) displays the resistivity data of three as-grown crystals before and after annealing for different annealing periods. While $\rho(T)$ curves obtained before annealing are similar for all the crystals, the curves determined after annealing show a significant increase in the $\rho(T)$ values and strong dependence on the annealing-time. The activation energy ($\Delta$), estimated from the fit to $\rho(T) = \rho_0 \exp(\Delta/T)$ near 300 K, shows similar trends. Quantitatively, the relative increases of $\rho(300 \text{ K})$ (and $\Delta$) after annealing are 450 (29%), 1609 (93%), and 284 (19%) times for annealing periods of 2, 8, and 14 days, respectively. Thus, the 8-day annealing period results in the largest increase in $\rho(300 \text{ K})$ and $\Delta$. In a previous study, $Ba_2Zn_2Fe_{12}O_{22}$ crystals grown with $BaO/B_2O_3$ flux showed an increase in the resistivity with $O_2$ annealing through the reduction in the oxygen deficiency, while crystals grown with $Na_2O/Fe_2O_3$ flux did not show any improvement in the resistivity with $O_2$ annealing, possibly owing to the contamination from Na impurities [7]. Therefore, in the case of our crystals grown from $Na_2O/Fe_2O_3$ flux, we postulate that the annealing period of 8 days is effective in removing oxygen deficiencies before the counter effect from the Na impurities increase considerably.

Furthermore, we observe that controlling the cooling rate is crucial to obtain an increase in the resistivity (see, figure 1(b)). An increase in the cooling rate results in a reduction in $\rho(300 \text{ K})$ and $\Delta$. Recent band structure calculations for $Ba_{2-x}Sr_xZn_2Fe_{12}O_{22}$ have indicated that the fraction of Zn at the tetrahedral sites in the magnetic $L$ blocks, denoted by $\gamma$ (figure 2(a)), is a critical parameter controlling the electrical properties of this system [8, 9]; when $\gamma$ increases, the compound becomes more insulating. In our study, after the elimination of the oxygen deficiency in the crystal by annealing under $O_2$ gas, the cooling rate is likely to control the distribution of Zn ions, i.e., $\gamma$. At 900°C, $\gamma$ is presumably close to a random distribution value of 0.5, and *quenching* will enable the crystal to maintain the high temperature $\gamma$ value down to room temperature. For the *slow* cooling condition, the crystal will have an equilibrium $\gamma$ value of temperatures lower than 900°C, which seems to be larger than 0.5 in this material system. This is consistent with the reported value of $\gamma$ (0.661) in $Ba_{0.5}Sr_{1.5}Zn_2Fe_{12}O_{22}$ single crystals [10, 11] grown under *slow* cooling condition [10]. From these results, we conclude that the combination of the 8-day $O_2$ annealing and the *slow* cooling is close to the optimal condition for obtaining the highest resistivity at 300 K (*most-insulating* sample).

The high resistivity of the *most-insulating* sample enabled us to detect the key features of the $H$-induced FE phase boundary, i.e., peaks in the $\varepsilon(H)$ data, even above room temperature as plotted in figure 3(a) and (b). Two main peaks in the $\Delta\varepsilon(H)/\varepsilon(0)$ curve (black ticks) clearly exist from 10 K to 310 K and then suddenly disappear at 315 K, indicative of ferroelectricity onset, in the *most-insulating* sample. The continuous evolution of these peak fields from low to high temperature regions supports that the *most-insulating* sample has indeed the $H$-induced FE transitions up to the 310 K. The corresponding $\varepsilon_2$ vs $H$ curves are also plotted at 10 and 300 K in figure 3(c). At the two critical magnetic fields for the ferroelectricity onset, there exist sharp peaks or dips in the $\varepsilon_2(H)$ curves. Particularly at 300 K, while a gradual increase of $\varepsilon_2$ exists in the background possibly due to increased leakage with $H$, we find dips at the expected FE phase transition fields. For this *most-insulating* sample, the loss tangent, $tan\ \delta \equiv \varepsilon_2/\varepsilon$ was about 1 at 300 K. In contrast, a typical as-grown sample had larger value of $tan\ \delta \cong 10$ at 300 K, and it did not show any dip or peak feature in the $\varepsilon(H)$ and $\varepsilon_2(H)$ curves above 130 K at the expected FE phase transition fields, presumably due to the increased leakage.

To directly confirm the development of ferroelectric polarization at the $\varepsilon$-peak-fields, we have tried to measure the magnetoelectric current $J_{ME}$. However, as we have prepared a relatively small piece of crystal for homogeneous heat treatments, we could not easily measure $J_{ME}$ reliably in the heat-treated sample in general. As a result, in the *most-insulating* sample, we could only confirm the existence of $P$ below 30 K through the peaks of $J_{ME}$, which almost match with the corresponding $\varepsilon(H)$ peaks. However, as the sample treated with 14 day-annealing and *slow* cooling had a little larger surface area, we could observe the clear peak features in the $J_{ME}$ curves up to 70 K. Figure 3 (d) summarizes the $P(H)$ curves of this sample at 10 K and 70 K, derived from the $J_{ME}$ measurements. Moreover, we could verify that the $H$ positions of the $J_{ME}$ peaks were matching well with those of $\varepsilon(H)$ peaks below 70 K. All these observations suggest that the two $\varepsilon(H)$ peaks observed up to 310 K in the *most-insulating* sample (figure 3(a) and (b)) indeed represent the evidence of the FE phase transitions.

The FE phase diagram thus determined from the two $\varepsilon(H)$ peaks is presented for the *most-insulating* in figure 4(a). Also plotted is the FE phase diagram of an as-grown crystal, for which $\varepsilon(H)$ peaks can be identified up to 220 K. The comparison shows that the FE phase boundaries of the *most-insulating* expand significantly to both low- and high-$H$ regions. In particular, the FE phase boundary in the low-field region moves abruptly toward zero $H$ around 280 K and remains there up to 310 K until it disappears above 315 K. As seen in figure 5(a), at the temperature of 315 K, both the $M/H$ curves measured at $\mu_0 H=0.01$ and 0.3 T start to deviate from the Curie-Weiss behavior and the $M/H$ curve measured at $\mu_0 H=0.01$ T shows a small kink. Thus, the helical spin ordering temperature $T_N$ of the *most-insulating* is considered to be 315 K. The disappearance of the FE phase above 315 K is then consistent with the expectation that the $H$-induced FE phase can be only realized below the spin ordering temperature.

It is a quite unexpected finding that the FE phase boundary in the *most-insulating* reaches vanishingly small magnetic fields above 270 K. According to the previous study in Ref. [5], the magnetic phase boundary of intermediate-III (solid lines) indeed coincides with the FE phase below 130 K. Furthermore, when the phase boundary of the intermediate-III is extended up to 300 K, the expected FE phase, if it still coincides with the intermediate-III phase, should be located between 1 T and 1.5 T around 300 K as drawn as solid lines in figure 4(a). To check the evolution of intermediate-III phase in the *most-insulating* and as-grown samples as well, we have measured the $M(H)$ curves at various temperatures. In figure 5(b), the $M(H)$ curves of the *most-insulating* are plotted for two representative temperatures, 150 K and 260 K. Following the method described in a previous study [5], the field at which the intermediate-III phase appears (solid triangle in the left) can be identified by a change in the slope of the $M(H)$ curve. Moreover, the intermediate-III phase disappears at the field where $M$ reaches the saturation value (solid triangle in the right). It is found that the black tick-position in the $\Delta\varepsilon(H)/\varepsilon(0)$ curve at 260 K, indicative of the low-$H$ FE phase boundary, is clearly different from the position of the solid triangle in the *most-insulating* crystal.

The magnetic phase diagrams thus summarized in figure 4(b) demonstrate that the phase boundaries of intermediate-III in the *most-insulating* sample clearly deviate from its FE ones shown in figure 4(a), particularly above 220 K and in the low-$H$ regions. For example, the FE phase boundaries at 300 K starts at almost zero magnetic fields while those of intermediate-III still remain above 0.7 T at 300 K. The deviation becomes more significant as temperature approaches the $T_N$. We note, however, that below 220 K, both FE and intermediate-III phases almost coincide each other in the *most-insulating* crystal. Even in the as-grown crystal, the phase boundaries of intermediate-III almost match with the FE ones below 220 K although they are narrower than those in Ref. [5] (solid lines). Therefore, the electric phase boundaries closely match with magnetic (intermediate-III) ones below 220 K, regardless of the heat treatment. This is also consistent with the results in Ref. [5], in which the FE and magnetic phase boundaries mostly match each other below 130 K. All these observations consistently suggest that the FE phase-coinciding with the intermediate-III phase at low temperatures-becomes independent of the intermediate-III phase, particularly near the spin ordering temperature and that it is a generic property of this hexaferrite system.

To explain the difference between the phase boundaries of ferroelectricity and intermediate-III observed in the *most-insulating* crystal, we employ the spin-current model [12], which predicts the direction of $P$ to be along $e \times Q$. Here, $e$ is the spin rotation axis and $Q$ is the propagation vector of the spiral ordering. $Q$ in $Ba_{0.5}Sr_{1.5}Zn_2Fe_{12}O_{22}$ is known to be parallel to the $c$-axis [10, 11]. To generate a nontrivial $P$ in the $ab$-plane, $e$ for the helimagnetic spin ordering should have a net component perpendicular to the $c$-axis. Then, such a spin ordering pattern is likely to have a canted conical spin structure with a cone axis that tilts off the $c$-axis, as shown in figure 2(b). It is noted that the suggested pattern is similar to the one recently proposed for $Ba_2Mg_2Fe_{12}O_{22}$ [13, 14]. Therefore, the appearance of the FE phase in $Ba_{0.5}Sr_{1.5}Zn_2Fe_{12}O_{22}$ can also be associated with the canted conical spin configuration, similar to the case of $Ba_2Mg_2Fe_{12}O_{22}$. At temperatures below 220 K, the intermediate-III phase seems to almost coincide with this canted conical phase in the *most-insulating* crystal. However, as the temperature increases, the spin anisotropy required to confine the spin-moments within the $ab$-plane will be naturally weakened. In such a situation, it is highly likely that the canted conical spin structure starts to appear at a much lower $H$ than the intermediate-III phase, explaining why the FE phase boundary in the *most-insulating* crystal moves to zero magnetic field near $T_N$. Therefore, we suggest that the canted conical spin state is uniquely realized in $Ba_{0.5}Sr_{1.5}Zn_2Fe_{12}O_{22}$ between 280 K and 315 K under a rather small $H$ value of 0.1 T. In particular, the temperature region to stabilize the canted conical structure is much higher for this compound than for $Ba_2Mg_2Fe_{12}O_{22}$, in which such a similar magnetic phase can only exist below its helical ordering temperature of 195 K [13].

4. **Conclusions**

We have observed the stabilization of the FE phase under a low magnetic field of 0.1 T

up to the temperature of 310 K in optimally heat-treated $Ba_{0.5}Sr_{1.5}Zn_2Fe_{12}O_{22}$. As a consequence, there appears a magnetodielectric effect of 1.1% under an $H$ value of 0.1 T at 300 K. The remarkable lowering of the critical magnetic field for switching ferroelectricity is understood as the result of a reduction in the spin anisotropy near $T_N$. The same mechanism can indeed be applied to a broader class of helimagnet and may further help to find new magnetoelectric phenomena in low magnetic-field and high temperature environment.

**Acknowledgments**

This work was financially supported by the NRL program (M10600000238), the GPP program of MEST (K20702020014-07E0200-01410), and the Fundamental R&D Program for Core Technology of Materials of MOKE. S. H. Chun was supported by the Seoul R&BD Program.

**Figure captions**

**Figure 1.** (a) Arrhenius plots of the resistivity $\rho$ in the *ab*-plane in the crystals heat-treated for different periods of time. All the crystals were cooled under the same *slow* condition, explained in the text. (b) The resistivity plots for the crystals cooled at different rates after annealing 8 days at 900°C.

**Figure 2.** (a) Schematic crystal structure of $Ba_{0.5}Sr_{1.5}Zn_2Fe_{12}O_{22}$ when viewed along the [110] direction. The helimagnetic structure discussed in the literature consists of alternating *L*- and *S*- spin blocks having large and small magnetic moments, respectively, lying in the *ab*-plane. (b) Suggested magnetic structure under a finite *H* in the FE phase of the *most-insulating* crystal. The structure can have conical spin ordering along the *c*-axis. Dashed line represents a spin cone axis deviating from the *c*-axis. The long and short arrows represent magnetic moments on the *L*-block and on the *S*-block, respectively.

**Figure 3.** $\Delta\varepsilon(H)/\varepsilon(0) \equiv (\varepsilon(H) - \varepsilon(0))/\varepsilon(0)$ plots of the *most-insulating* (a) below 220 K and (b) above 260 K. In (b), each curve is shifted by an arbitrary offset for clarity. (c) $\varepsilon_2$ vs. *H* curves of the *most-insulating* crystal at 10 K and 300 K. (d) *P* vs *H* curves at 10 and 70 K for the sample subject to the 14 day-annealing and *slow* cooling treatments.

**Figure 4.** (a) The FE phase diagrams of the as-grown and *most-insulating* crystals derived from the peak positions of $\varepsilon(H)$ data. The intermediate-III (magnetic) phase boundary (solid lines) in Ref. [5] is also plotted for comparison. The dashed line refer to the temperature of $T_N$. (b) Intermediate-III phase diagrams of the as-grown and *most-insulating* crystals, derived from the *M(H)* measurements as illustrated in figure 5(b). The intermediate-III phase boundary in Ref. [5]

is plotted again for comparison.

**Figure 5.** (a) *M/H* vs. *T* curves for the *most-insulating* crystal measured at $\mu_0H$=0.01 and 0.3 T after field cooling. (b) *M* vs. *H* curve for the *most-insulating* crystal at 150 and 260 K. The red dashed lines are the linear guide for the eye. Black ticks represent the magnetic fields at which the $\varepsilon$ maximum, thus ferroelectricity, develops. Solid triangles represent two magnetic fields between which the intermediate-III phase is stabilized (see text for details).

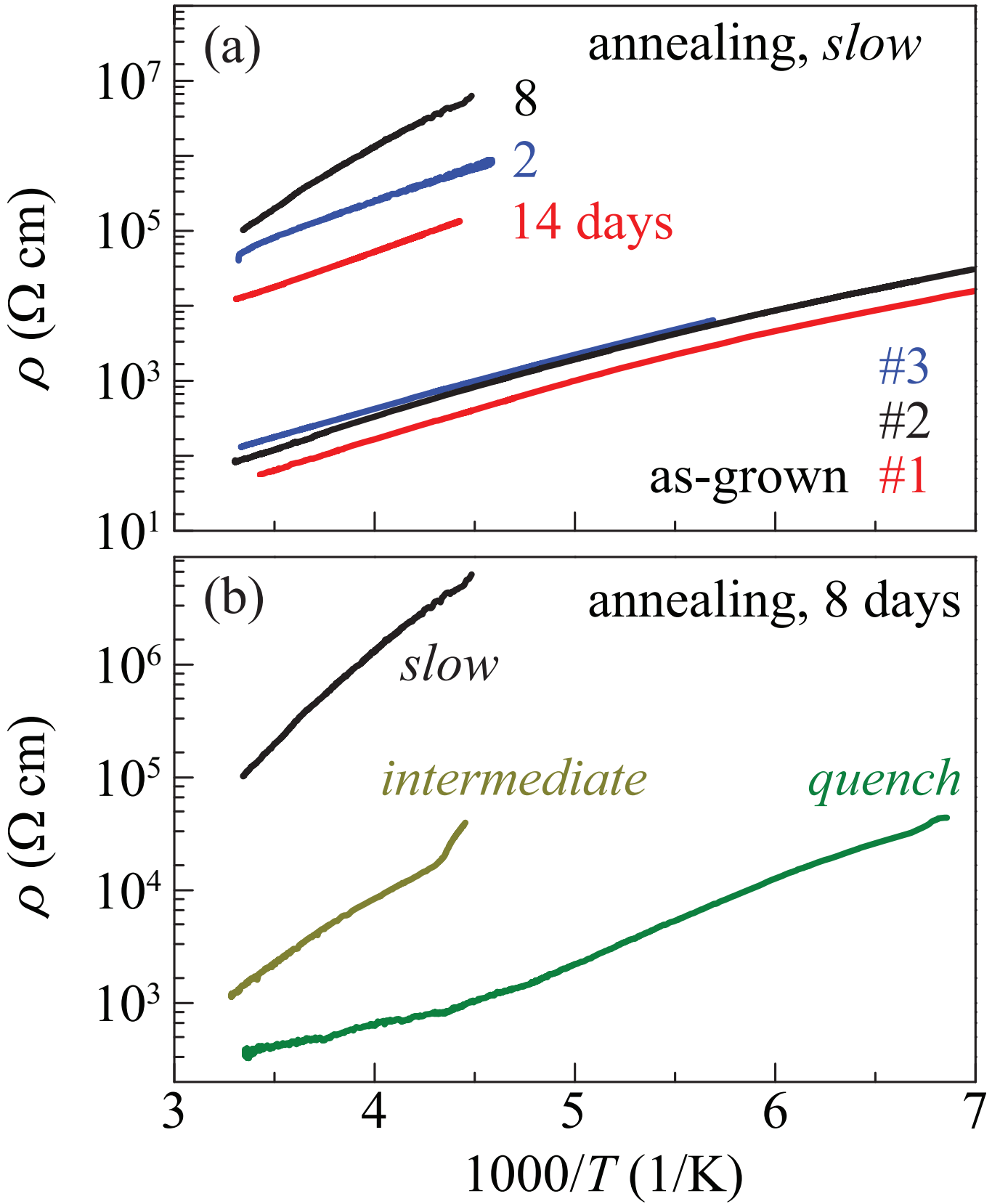

Figure 1 (figure1.eps)

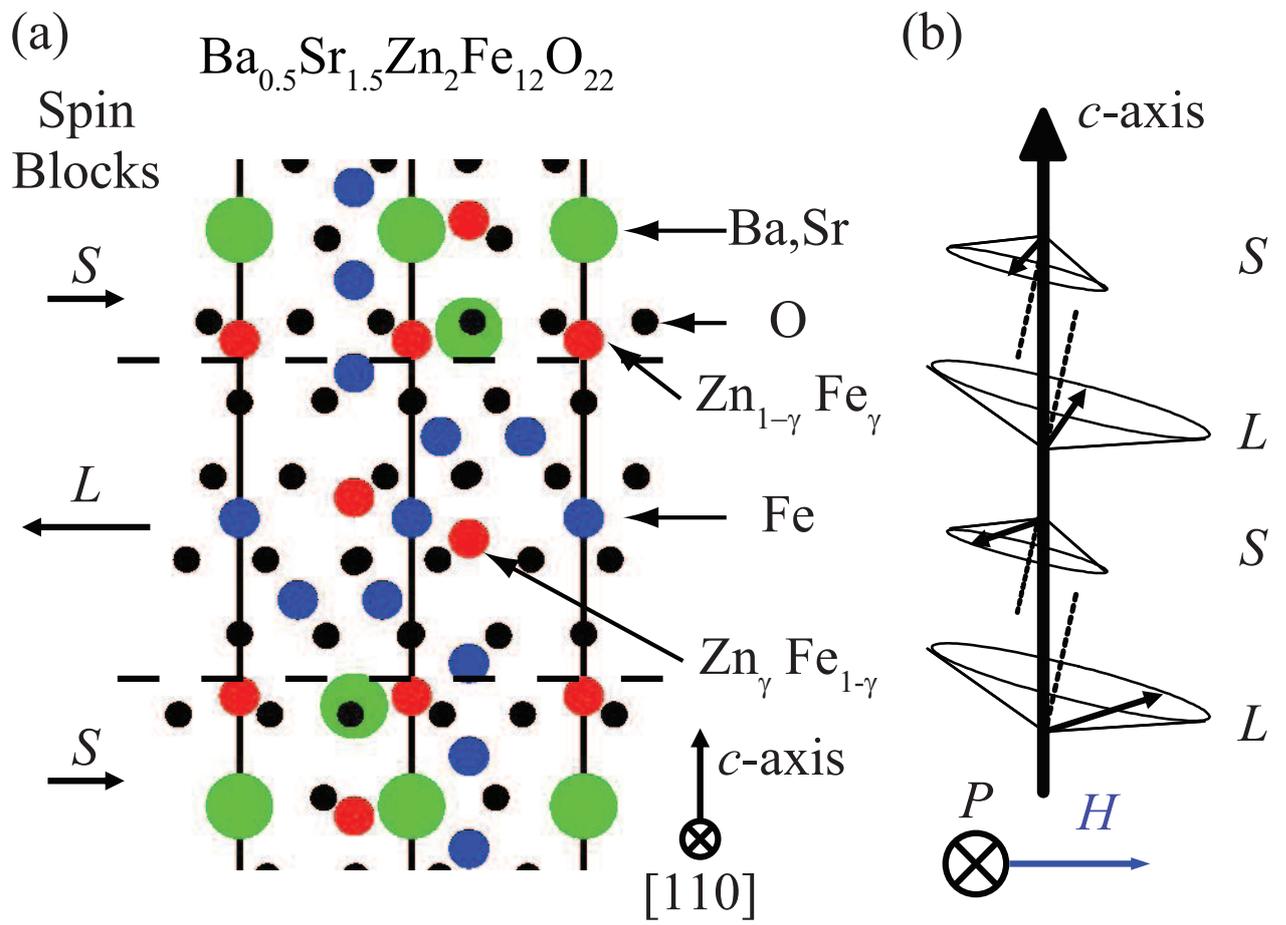

Figure 2 (figure2.eps)

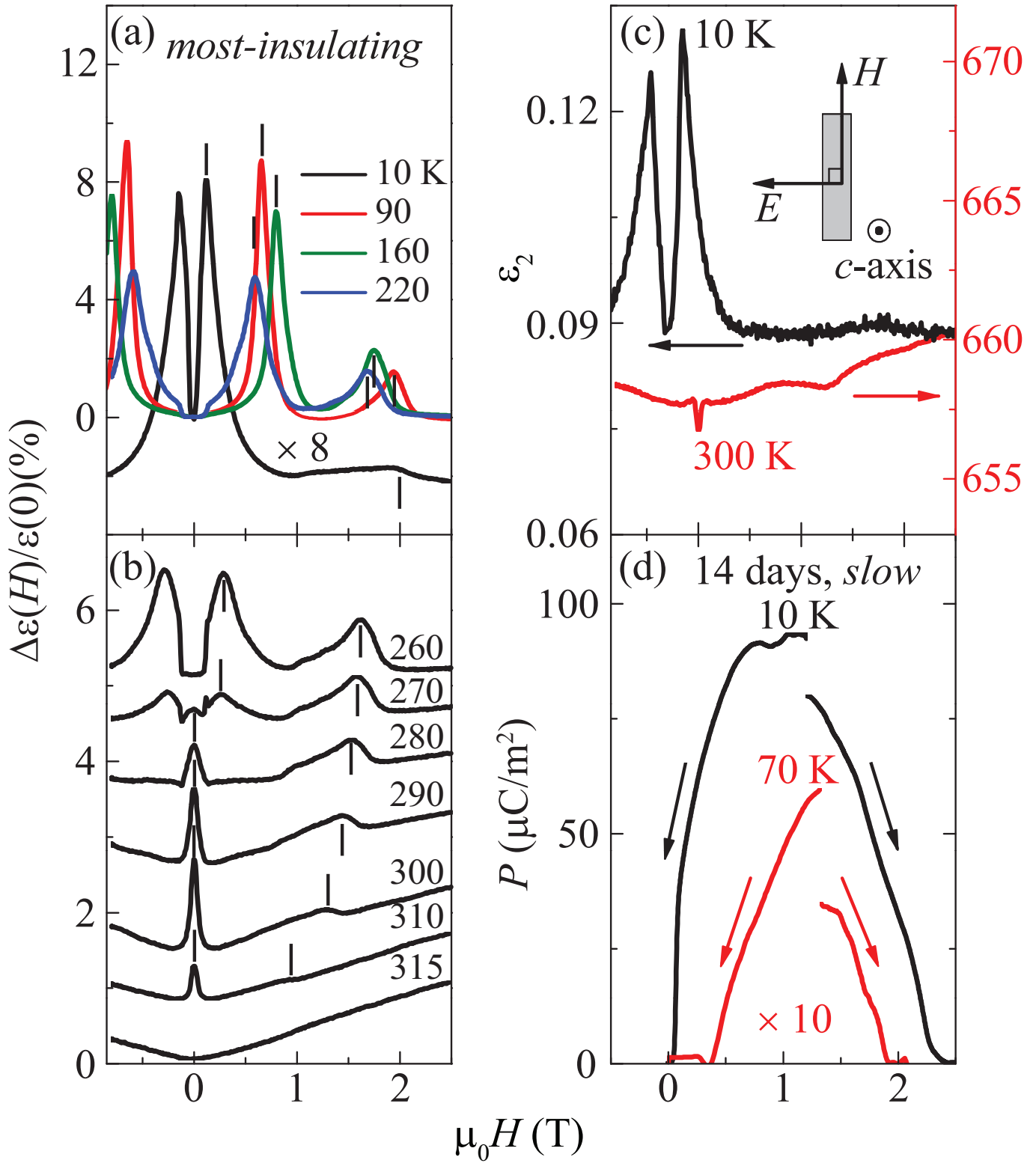

Figure 3 (figure3.eps)

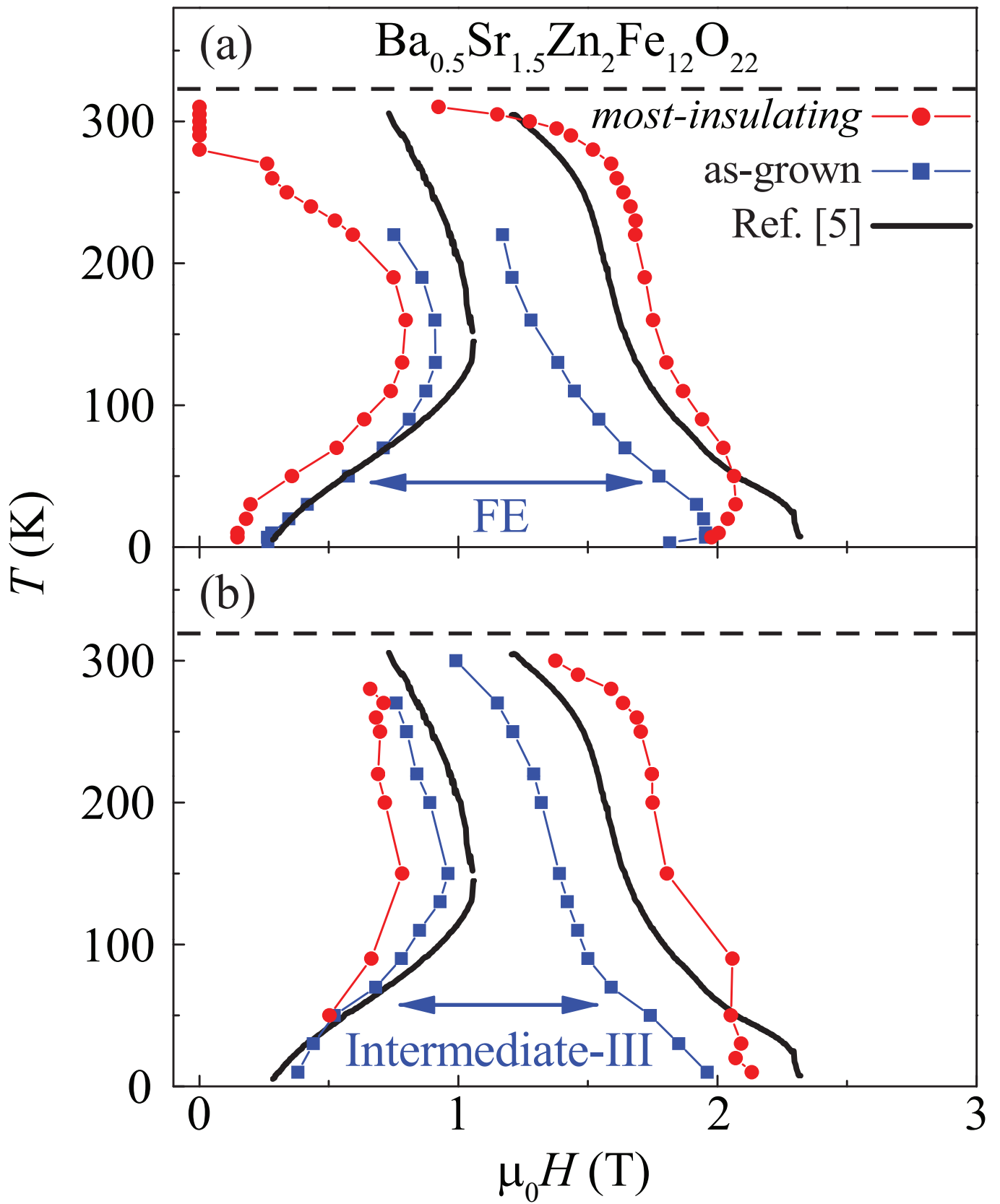

Figure 4 (figure4.eps)

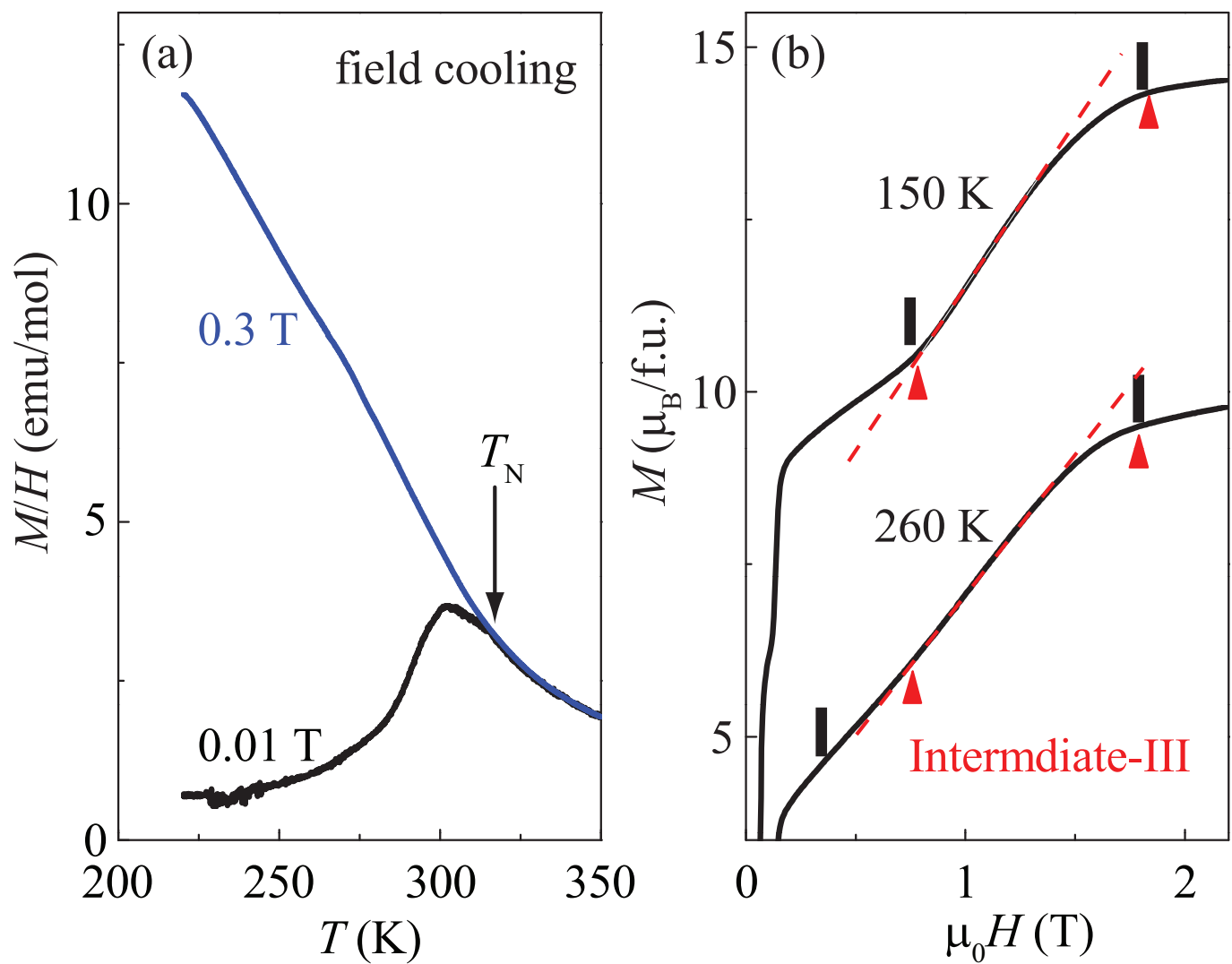

Figure 5 (figure5.eps)